# Expression of Generalized Newton Iteration Method via Generalized Local Fractional Taylor Series

**Yang Xiao-Jun**

Department of Mathematics and Mechanics, China University of Mining and Technology, Xuzhou Campus, Xuzhou, Jiangsu,221008, P. R. China

Email: dyangxiaojun@163.com

**Abstract** –Local fractional derivative and integrals are revealed as one of useful tools to deal with everywhere continuous but nowhere differentiable functions in fractal areas ranging from fundamental science to engineering. In this paper, a generalized Newton iteration method derived from the generalized local fractional Taylor series with the local fractional derivatives is reviewed. Operators on real line numbers on a fractal space are induced from Cantor set to fractional set. Existence for a generalized fixed point on generalized metric spaces may take place.

## 1. Introduction

Over the past ten years, local fractional calculus [1-8] played an important part in fractal mathematics and engineering, especially in nonlinear phenomena. More recently, Newton iteration method based on local fractional calculus was presented in [6]. A procedure for the Newton iteration method is not easy enough for an engineer to master it. In order to express the generalized Newton iteration method, this paper proposes operators on real line numbers on a fractal space. Taking the definitions of local fractional derivatives and local fractional integrals into account and using generalized local fractional Taylor series [11, 15, 16], it is easy to understand the generalized Newton iteration method.

This paper is organized as follows. In section 2, operators on real line numbers on a fractal space are investigated. Generalized local fractional Taylor series are derived from local fractional calculus in section 3. Generalized Newton iteration method is discussed in section 4. The conclusions are in section 5.

## 2. Operators on real line numbers on a fractal space

At first, we start with operator on real line numbers in fractal space.

**Definition 1** Let $\mathbb{R}$ be real number. There exist responding real line numbers on a fractal set $E$ with fractional dimension $\alpha$, noted by $\mathbb{R}_1^\alpha$.

**Definition 2** A metric for a fractal set $E$ with fractional dimension $\alpha$, $0 < \alpha \leq 1$, is a map $\rho_\alpha : E \times E \to \mathbb{R}_1^\alpha$ such that for all $x^\alpha, y^\alpha, z^\alpha \in E$.

(1) $\rho_\alpha(x^\alpha, y^\alpha) \geq 0$ with equality if $x^\alpha = y^\alpha$;

(2) $\rho_\alpha(x^\alpha, y^\alpha) = \rho_\alpha(y^\alpha, x^\alpha)$;

(3) $\rho_\alpha(x^\alpha, z^\alpha) \leq \rho_\alpha(x^\alpha, y^\alpha) + \rho_\alpha(y^\alpha, z^\alpha)$.

The pair $(E, \rho_\alpha)$ is a generalized metric space.

**Remark.1** $\mathbb{R} = \{$numbers on a fractal set $E$ with $\alpha = 1\}$.

**Remark. 2** Let $E = \mathbb{R}_n^\alpha$.

$$\rho_{\alpha,2}(x^\alpha, y^\alpha) = \left(\sum_{i=1}^n (x_i - y_i)^{2\alpha}\right)^{\frac{1}{2}}$$

is a metric on $\mathbb{R}_n^\alpha$, where

$$x^\alpha = (x_1^\alpha, x_2^\alpha, ..., x_n^\alpha) \in \mathbb{R}_n^\alpha$$

and

$$y^\alpha = (y_1^\alpha, y_2^\alpha, ..., y_n^\alpha) \in \mathbb{R}_n^\alpha.$$

Similarly, if $E = \mathfrak{R}_n^\alpha = \mathbb{R}_n^\alpha \times \mathbb{R}_n^\alpha$, we have

$$\rho_{\alpha,2}(z_1^\alpha, z_2^\alpha) = \left(\sum_{i=1}^n (z_{1,i} - z_{2,i})^{2\alpha}\right)^{\frac{1}{2}}$$

is a metric on complex space $\mathfrak{R}_n^\alpha$, where

$$z_1^\alpha = (z_{1,1}^\alpha, z_{1,2}^\alpha, ..., z_{1,n}^\alpha) \in \mathfrak{R}_n^\alpha \text{ and}$$

$$z_2^\alpha = (z_{2,1}^\alpha, z_{2,2}^\alpha, ..., z_{2,n}^\alpha) \in \mathfrak{R}_n^\alpha.$$

There is a simple example that for $z^\alpha = x^\alpha + i^\alpha y^\alpha$

$|z^\alpha| = \sqrt{x_1^{2\alpha} + x_2^{2\alpha}}$ is referred [9].

**Remark.3** In special case of $\alpha = 1$ it follows that both $\mathbb{R}_n^1 = \overbrace{\mathbb{R} \times \mathbb{R} \times ... \times \mathbb{R}}^{n}$ and $\mathbb{R}$ are classical complex spaces and that both $\mathfrak{R}_1^1 = \mathbb{R} \times \mathbb{R}$ and

$\mathfrak{R}_n^1 = \overbrace{\mathfrak{R}_1^1 \times \mathfrak{R}_1^1 \times ... \times \mathfrak{R}_1^1}^{n}$ are classical complex spaces.

If $a^\alpha, b^\alpha, c^\alpha$ belong to the set $\mathbb{R}_1^\alpha$ of real line numbers, then

(1) $a^\alpha + b^\alpha$ and $a^\alpha b^\alpha$ belong to the set $\mathbb{R}^\alpha$;

(2) $a^\alpha + b^\alpha = b^\alpha + a^\alpha = (a+b)^\alpha = (b+a)^\alpha$;

(3) $a^\alpha + (b^\alpha + c^\alpha) = (a^\alpha + b^\alpha) + c^\alpha$;

(4) $a^\alpha b^\alpha = b^\alpha a^\alpha = (ab)^\alpha = (ba)^\alpha$;

(5) $a^\alpha (b^\alpha c^\alpha) = (a^\alpha b^\alpha) c^\alpha$;

(6) $a^\alpha (b^\alpha + c^\alpha) = a^\alpha b^\alpha + a^\alpha c^\alpha$;

(7) $a^\alpha + 0^\alpha = 0^\alpha + a^\alpha = a^\alpha$

and $a^\alpha \cdot 1^\alpha = 1^\alpha \cdot a^\alpha = a^\alpha$.

The above is operators on real line numbers on a fractional space.

## 3. Generalized local fractional Taylor series

### 3.1 Review for local fractional derivatives and local fractional integrals

**Definition 3** A non-differentiable function
$$f: \mathbb{R} \to \mathbb{R}, \ x \to f(x)$$
is called to be local fractional continuous, when
$$|f(x) - f(x_0)| < \varepsilon^\alpha \quad (1)$$
with $|x - x_0| < \delta$, for $\varepsilon, \delta > 0$ and $\varepsilon, \delta \in \mathbb{R}$. If $f(x)$ is local fractional continuous on the interval $(a,b)$, we denote $f(x) \in C_\alpha(a,b)$.

**Definition 4** The local fractional derivative of $f(x)$ of order $\alpha$ at $x = x_0$ is defined [6-9]
$$f^{(\alpha)}(x_0) = \frac{d^\alpha f(x)}{dx^\alpha}\bigg|_{x=x_0} = \lim_{x \to x_0} \frac{\Delta^\alpha(f(x) - f(x_0))}{(x - x_0)^\alpha}, \quad (2)$$
where $\Delta^\alpha(f(x) - f(x_0)) \cong \Gamma(1+\alpha)\Delta(f(x) - f(x_0))$.

For any $x \in (a,b)$, there exists
$$f^{(\alpha)}(x) = D_x^{(\alpha)} f(x),$$
denoted by
$$f(x) \in D_x^{(\alpha)}(a,b).$$

**Definition 5** Local fractional integral of $f(x)$ of order $\alpha$ in the interval $[a,b]$ is defined [6-9, 12-16]
$$\begin{aligned}{}_aI_b^{(\alpha)} f(x) &= \frac{1}{\Gamma(1+\alpha)} \int_a^b f(t)(dt)^\alpha \\ &= \frac{1}{\Gamma(1+\alpha)} \lim_{\Delta t \to 0} \sum_{j=0}^{j=N} f(t_j)(\Delta t_j)^\alpha\end{aligned} \quad (3)$$

where $\Delta t_j = t_{j+1} - t_j$, $\Delta t = \max\{\Delta t_1, \Delta t_2, \Delta t_j, ...\}$ and $[t_j, t_{j+1}]$, $j = 0, ..., N-1$, $t_0 = a, t_N = b$, is a partition of the interval $[a,b]$.

Here, it follows that
$${}_aI_a^{(\alpha)} f(x) = 0 \text{ if } a = b$$
and
$${}_aI_b^{(\alpha)} f(x) = -{}_bI_a^{(\alpha)} f(x) \text{ if } a < b.$$

For any $x \in (a,b)$, there exists ${}_aI_x^{(\alpha)} f(x)$, denoted by $f(x) \in I_x^{(\alpha)}(a,b)$.

**Remark 4.** If $f(x) \in D_x^{(\alpha)}(a,b), {}_{x_0}I_x^{(\alpha)}(a,b)$, we have $f(x) \in C_\alpha(a,b)$.

**Remark 5.** Above definition is deferent from Jumarie [10] and Kolwankar [1,4]. Meanwhile, $\alpha$ is fractal dimension and there exist the following relations
$$a^\alpha + b^\alpha = (a+b)^\alpha$$
and
$$a^\alpha - b^\alpha = (a-b)^\alpha.$$

In order to take the above into account, we have the equality
$$\frac{1}{\Gamma(1+\alpha)} \int_x^y (dt)^\alpha = \frac{1}{\Gamma(1+\alpha)} \int_c^y (dt)^\alpha + \frac{1}{\Gamma(1+\alpha)} \int_x^c (dt)^\alpha,$$
which provides
$$\frac{1}{\Gamma(1+\alpha)}(y-x)^\alpha = \frac{1}{\Gamma(1+\alpha)}(y-c)^\alpha + \frac{1}{\Gamma(1+\alpha)}(c-x)^\alpha.$$

In this case we have the following relation
$$(y-x)^\alpha = (y-c)^\alpha + (c-x)^\alpha.$$

Furthermore,
$$(y+x)^\alpha = (y-c)^\alpha + (c+x)^\alpha.$$

When $c = 0$, we can find that
$$(y+x)^\alpha = y^\alpha + x^\alpha, \quad (4)$$

It follows from (4) that
$$(y+x)^\alpha - y^\alpha = x^\alpha = (y+x-y)^\alpha,$$
Hence
$$(x \pm y)^\alpha = x^\alpha \pm y^\alpha. \quad (5)$$

### 3.2 Generalized local fractional Taylor series

**Lemma 1** (Generalized local fractional Taylor theorem) Suppose that $f^{((k+1)\alpha)}(x) \in C_\alpha(a,b)$, for $k = 0, 1, ..., n$, $0 < \alpha \leq 1$, then we have[11]
$$f(x) = \sum_{k=0}^{n} \frac{f^{(k\alpha)}(x_0)}{\Gamma(1+k\alpha)}(x-x_0)^{k\alpha} + \frac{f^{((n+1)\alpha)}(\xi)}{\Gamma(1+(n+1)\alpha)}(x-x_0)^{(n+1)\alpha} \quad (6)$$

with $a < x_0 < \xi < x < b$, $\forall x \in (a,b)$, where
$$f^{((k+1)\alpha)}(x) = \overbrace{D_x^{(\alpha)}...D_x^{(\alpha)}}^{k+1 \text{ times}} f(x).$$

**Theorem 2** (Generalized local fractional Taylor series)

Suppose that $f^{((k+1)\alpha)}(x) \in C_\alpha(a,b)$, for $k=0,1,...,n$, $0<\alpha\leq 1$, then we have

$$f(x) = \sum_{k=0}^{\infty} \frac{f^{(k\alpha)}(x_0)}{\Gamma(1+k\alpha)}(x-x_0)^{k\alpha}, 0<\alpha\leq 1 \qquad (7)$$

with $a < x_0 < \xi < x < b$, $\forall x \in (a,b)$,

where $f^{((k+1)\alpha)}(x) = \overbrace{D_x^{(\alpha)}...D_x^{(\alpha)}}^{k+1\ times} f(x)$.

*Proof.* From (6), taking the limit $n \to \infty$, we have the result.

As a direct result, we have the local fractional MC-Laurin's series, which is read as

$$f(x) = \sum_{k=0}^{\infty} \frac{f^{(k\alpha)}(0)}{\Gamma(1+\alpha k)} x^{k\alpha}, 0<\alpha\leq 1. \qquad (8)$$

## 4. Generalized Newton iteration method

In this section, an iteration method via the local fractional derivative for a non-differential function is called generalized local fractional iteration method. This method is to solve the local fractional equation.

Suppose that $\varphi: \mathbb{R} \to \mathbb{R}$, $x^\alpha \to \varphi_\alpha(x)$, is a $\alpha th$ continuously non-differentiable function, for $0<\alpha\leq 1$. There exists $\xi \in [x^*, x_{k-1}]$ such that

$$\begin{aligned}|x^{*\alpha} - x_k^\alpha| &= |\varphi_\alpha(x^*) - \varphi_\alpha(x_{k-1})| \\ &= \left|\frac{\varphi^{(\alpha)}(\xi)}{\Gamma(1+\alpha)}\right|\left|(x^* - x_{k-1})^\alpha\right| \leq L|x^{*\alpha} - x_{k-1}^\alpha|\end{aligned} \qquad (9)$$

We have the inequality

$$|x^{*\alpha} - x_k^\alpha| \leq L|x^* - x_{k-1}|^\alpha \leq \cdots \leq L^k|x^* - x_0|^\alpha = L^k|x^{*\alpha} - x_0^\alpha|,$$

where $\left|\frac{\varphi^{(\alpha)}(x)}{\Gamma(1+\alpha)}\right| \leq L < 1$ for any $x \in [a,b]$.

Therefore $x_k \to x^*$ as $k \to \infty$.

For any positive integer $p$ we can find that

$$\begin{aligned}|x_{k+p}^\alpha - x_k^\alpha| &\leq |x_{k+p}^\alpha - x_{k+p-1}^\alpha| \\ &\quad + |x_{k+p-1}^\alpha - x_{k+p-2}^\alpha| + \cdots + |x_{k+1}^\alpha - x_k^\alpha| \\ &\leq L^{p-1}|x_{k+1}^\alpha - x_k^\alpha| + L^{p-2}|x_{k+1}^\alpha - x_k^\alpha| \\ &\quad + \cdots + |x_{k+1}^\alpha - x_k^\alpha| = (L^{p-1} + L^{p-2} + \cdots + 1)|x_{k+1}^\alpha - x_k^\alpha|.\end{aligned} \qquad (10)$$

It follows from (10) that

$$|x^{*\alpha} - x_k^\alpha| \leq \frac{1}{1-L} < |x_{k+1}^\alpha - x_k^\alpha| \text{ as } p \to \infty,$$

Hence, there exists a local fractional iteration process

$$x_{k+1}^\alpha = \varphi_\alpha(x_k), \qquad (11)$$

which converges the root of $\phi_\alpha(x) = 0$.

Notice: Both $\phi_\alpha(x)$ and $\varphi_\alpha(x)$ are local fractional continuous functions. Hence, from (12) we may obtain a new fixed point.

## 5 Remarking conclusions

In the present paper, a new iteration method via the local fractional Taylor series with local fractional derivatives, a generalized local fractional iteration method, is studied. The method is to solve the equations of non-differential functions defined on fractal sets. Real line numbers on fractal sets are investigated, and it may be a new subject on fractal geometrics. As classical iteration method, there may be a new fixed point method in a fractal space. It is to say, the existence for a generalized fixed point on generalized metric spaces [15-16] may take place.